\documentclass[10pt,conference]{IEEEtran}
\IEEEoverridecommandlockouts
\usepackage{array}
\usepackage[utf8]{inputenc} 
\usepackage[T1]{fontenc}    
\usepackage{url}            
\usepackage{booktabs}       
\usepackage{amsfonts}        
\usepackage{microtype}      
\usepackage{graphicx}
\usepackage{color,ulem}
\usepackage{amsthm,amsmath,bbm}
\usepackage{flushend}
\usepackage{listings}
\usepackage{verbatim}
\usepackage{mathtools}
\usepackage{pythonhighlight}
\usepackage{adjustbox}

\newcommand{\BOBO}{\texttt{ibmq\_boeblingen}}
\newcommand{\JOJO}{\texttt{ibmq\_johannesburg}}

\newcommand{\VAL}{\texttt{ibmq\_valencia}}
\newcommand{\CAM}{\texttt{ibmq\_cambridge}}

\newcommand{\YORK}{\texttt{ibmq\_5\_yorktown}}

\newcommand{\metricId}{$\Vert (\Tilde{K} - \mathbbm{1}) \Vert_F$}
\newcommand{\metricK}{$\Vert (\Tilde{K} - K) \Vert_F$}
\newcommand{\metricF}{$\Delta(f)$}

\newcommand{\tabhead}{Construction $\Tilde{K}$ & \metricId &  \metricK & \metricF  \\ \hline}

\newcommand{\qkern}[1]{K_{\{#1\}}}

\begin{document}

\title{Scalable quantum processor noise characterization\thanks{This manuscript has been authored by UT-Battelle, LLC, under Contract No. DE-AC0500OR22725 with the U.S. Department of Energy. The United States Government retains and the publisher, by accepting the article for publication, acknowledges that the United States Government retains a non-exclusive, paid-up, irrevocable, world-wide license to publish or reproduce the published form of this manuscript, or allow others to do so, for the United States Government purposes. The Department of Energy will provide public access to these results of federally sponsored research in accordance with the DOE Public Access Plan.}}

\author{
  Kathleen E.~Hamilton ~\IEEEmembership{}
  Tyler~Kharazi ~\IEEEmembership{}
  Titus~Morris ~\IEEEmembership{}\\
  Alexander J. McCaskey ~\IEEEmembership{}
  Ryan S. Bennink ~\IEEEmembership{}
  and Raphael C.~Pooser ~\IEEEmembership{}%
\IEEEcompsocitemizethanks{\IEEEcompsocthanksitem K. Hamilton, T. Kharazi, R. Bennink, and R. Pooser are with the Computer Science and Engineering Division, Oak Ridge National Laboratory, Oak Ridge, Tennessee .\protect\\
E-mail: hamiltonke@ornl.gov
\IEEEcompsocthanksitem T. Morris is with the Physical Sciences Division, Oak Ridge National Laboratory Oak Ridge, Tennessee.%
\IEEEcompsocthanksitem A. McCaskey is with the Computer Science and Mathematics Division, Oak Ridge National Laboratory, Oak Ridge, Tennessee.}%
\thanks{Manuscript received XXXX; revised YYYY.}}

\maketitle

\begin{abstract}
Measurement fidelity matrices (MFMs) (also called error kernels) are a natural way to characterize state preparation and measurement errors in near-term quantum hardware.  They can be employed in post processing to mitigate errors and substantially increase the effective accuracy of quantum hardware.  However, the feasibility of using MFMs is currently limited as the experimental cost of determining the MFM for a device grows exponentially with the number of qubits. In this work we present a scalable way to construct approximate MFMs for many-qubit devices based on cumulant expansion. Our method can also be used to characterize various types of correlation error.
\end{abstract}

\section{Introduction}
\label{sec:introduction}
The current era of quantum computing has been characterized as the ``noisy intermediate scale quantum '' (NISQ) era \cite{preskill2018quantum}.  While fault tolerance and error corrected qubits are necessary for many large-scale quantum algorithms,  recent studies have suggested that quantum processors with 50-60 qubits and sufficiently low error rates can out-perform classical computers at certain problems \cite{arute2019quantum,bremner_achieving_2017,aaronson_complexity-theoretic_2016}. Devices with several 10's of qubits are now publicly accessible and allow diverse end users to explore a variety of applications. There is a need for benchmarks and metrics that can be executed by both hardware developers end users to quantify device errors and application performance \cite{michielsen2017benchmarking,wright2019benchmarking}.  As the size of available devices and the applications deployed continue to grow, the benchmarks employed will need to scale as well.   

Quantum processors exhibit multiple error types, including state preparation and measurement (SPAM) error, gate errors, and cross-talk. These errors have motivated the development of  error characterization and mitigation methods, each of which pose unique challenges in scalability.  Extrapolating zero-noise behavior can be difficult for densely parameterized circuits \cite{temme2017error,li2017efficient,havlivcek2019supervised};  pulse-level control methods may not be available to end users \cite{Kandala2019}; and tomographic-based approaches have large computational overhead \cite{endo2018practical}. Error mitigation routines are available in the IBM Qiskit Ignis library \cite{Qiskit}. 

Matrix-based methods of noise characterization are commonly employed in quantum state tomography, where the process matrix and measurement fidelity matrix (MFM) are used to characterize gate fidelity \cite{PhysRevB.90.144504,bialczak2010quantum}.  Measurement errors have been commonly treated as independent  \cite{steffen2006measurement,kandala2017hardware}, with $n$-qubit matrices built from single qubit measurements \cite{bialczak2010quantum}.  In this work we present a scalable matrix-based characterization of SPAM error which uses MFMs to store the distribution of results obtained when each computational state is prepared and subsequently measured.  

MFMs capture the net effect of multiple sources of error in qubit initialization, gates, and measurement.  Recent studies have shown that MFMs can be used to characterize the individual sources of noise on hardware \cite{PhysRevX.9.021045}.  However a significant challenge to this approach is the exponential scaling of the number of experiments needed to determine a full $n$-qubit MFM.  While determining a MFM is significantly less costly than quantum tomography \cite{endo2018practical} or Richardson extrapolation~\cite{temme2017error}, to construct a $2^n \times 2^n$ MFM a minimum of $2^n$ different circuits must each be executed many times to accumulate significant statistics.  This is cost-prohibitive for remote users who can access the device only intermittently and for limited amounts at a time.  Previous studies have introduced the approach of constructing MFMs from single-qubit measurements \cite{bialczak2010quantum}. While such an approach is scalable, it cannot capture error correlations that are widely observed.

Here we present scalable methods for constructing  many-qubit MFMs and demonstrate these methods on quantum processors with superconducting qubits. These methods are based on the cumulant expansion which is commonly employed in the study of correlated systems across many physics disciplines \cite{kubocumulant}.  This approach provides a systematic method to incorporate correlations into the construction of an MFM and a tool to characterize correlated errors on quantum devices that are accessed via cloud-based queue systems.

In Section \ref{sec:methods} we describe the methods used to generate state preparation circuits and characterize correlated SPAM errors. In Sections  \ref{sec:cumulant_MFM} and \ref{sec:cluster_MFM} we apply our methods to multiple IBM qubit devices and present results showing how error correlations can be quantified by comparing the full MFM of a set of qubits to an approximation of the MFM constructed from measurements on smaller qubit subsystems.  In Section \ref{sec:discussion} we discuss these results, and present our conclusions in Section \ref{sec:conclusions}.

\section{Methods}
\label{sec:methods}
\subsection{Circuits for Measuring MFMs}
A circuit $\mathcal{U}$ is a sequence of one- and two-qubit gates that rotates an $n$-qubit register $|0\rangle^{\otimes n}$ to a final quantum state $|\psi \rangle$. This state is sampled (measured in the computational basis) yielding a string of $n$ bits. Preparing and sampling the state $n_s \gg 1$ times yields an estimate of the probability distribution $p$ over all $2^n$ possible bitstrings $x_i \in \{0,1\}^n$. 

We use circuits $\lbrace \mathcal{U}(x_i)\rbrace $that are constructed to return a specific basis state ($x_i$), such that in the absence of any hardware errors, sampling would return the bit string $x_i$ with probability 1.  When hardware errors occur, sampling the state $\mathcal{U}(x_i) |0\rangle^{\otimes n}$ may sometimes yield counts in states other than the target state $x_i$, i.e.  $p(x_i)\leq 1$ and $p(x_j)> 0$ for some $x_j \ne x_i$. The distribution of counts in non-target states is dependent on multiple sources of noise including initialization error, gate errors, and measurement errors. 

A MFM is a matrix $K$ whose entries $K_{ij} = p(x_j|x_i)$ are the conditional probabilities that sampling state $\mathcal{U}(x_i) |0\rangle^{\otimes n}$ yields $x_j$. The distribution $q$ of an arbitrary state $|\psi\rangle$ can be approximated as $q = K q_\text{ideal}$ where $q_\text{ideal}$ is the output distribution in the absence of errors. 

Many methods for constructing MFMs are being developed.  The most direct method is to construct each row $i$ of the matrix by repeatedly initializing an $n$-qubit register in the state $|0\rangle^{\otimes n}$, applying $X$ gates to individual qubits to nominally prepare the state $|x_i\rangle$ and measuring all $n$ qubits in the computational basis. This approach is not scalable: if there are $2^n$ possible target states, $2^n$ circuits must be run to estimate the full MFM.

In the remainder of the section we describe several methods to approximate many-qubit MFMs by combining MFMs of much smaller subsystems. We use $K$ to denote the full MFM of an $n$ qubit system, $K_{Q}$ to denote the MFM of qubit or subsystem $Q$, and $\tilde{K}$ to denote an approximation to $K$ obtained by combining subsystem MFMs.

\subsection{Cumulant expansion for MFMs}
\label{sec:cumulant_exp}
A cumulant expansion \cite{kubocumulant} relates the moments of a function to a set of generating coefficients called cumulants. If the moments of $n$ random variables are encoded as the coefficients of a multivariate polynomial $M(t_1,\ldots,t_n)$, the cumulants are the coefficients of the generating function $\lambda(t_1,\ldots,t_n)$ defined by
\begin{equation}
    M(t_1,\ldots,t_n) = e^{\lambda(t_1,\ldots,t_n)}.
\end{equation}
Each $m$th order moment is function of cumulants up to order $m$. The advantage of the cumulant expansion is that only a few cumulants are needed when correlations among all the variables are due primarily to the combined effect of few-variable correlations. Intuitively, an $m$th order cumulant quantifies the component of an $m$th order moment that is not due to a combination of lower-order moments.

The cumulant expansion can be applied to MFMs in a natural way.
An $n$-qubit MFM with elements $K_{ij} = p(x_j | x_i)$ can be viewed as a function of $2n$ variables. The low order (few-qubit) moments of $K$ can be used to calculate low order cumulants, which then generate a full conditional probability matrix $\tilde{K}$.
If $K$ is dominated by few-qubit correlations, the generated matrix $\tilde{K}$ will be close to $K$.  In this case $K$ may be efficiently estimated by a polynomial number of few-qubit circuits and measurements.  Furthermore, extrapolated MFMs may be compared to explicitly measured MFMs to characterize correlated errors in NISQ devices. While such characterization would be unscalable for arbitrarily high order correlations, modern devices tend to be limited in the degree of correlated noise between qubits by the connectivity, distance from one another, and the readout resonator configuration (in the case of superconducting devices). Therefore our method can be used to characterize correlated noise within small collections of qubits and within subregions of a chip.

For a circuit $\mathcal{U}(x_i) = u_0 \otimes u_1 \otimes \dots u_{n-1}$, sampling from the final prepared state returns $\langle u_0 u_1 \dots u_{n-1} \rangle$. In the absence of any correlated noise or error in the hardware then we could decompose the final distribution as $\langle U_0 U_1 \dots U_{n-1} \rangle =\langle U_0 \rangle \langle U_1\rangle \dots \langle U_{n-1}\rangle$ and these matrices could be constructed from the $n$ single qubit conditional probabilities.

We use $\lambda_{Q}$ to denote a cumulant coefficient involving a subset $Q$ of qubits.
The single-qubit cumulants and conditional probabilities for a qubit $a$ are the same:
\begin{equation}
\begin{aligned}
\lambda_{a}(0 | 0)&=p_{a}(0 | 0) & \lambda_{a}(1 | 0)&=p_{a}(1 | 0) \\
\lambda_{a}(0 | 1)&=p_{a}(0 | 1) & \lambda_{a}(1 | 1)&=p_{a}(1 | 1).
\end{aligned}
\label{eq:one_body_terms}
\end{equation}
Whereas the two-qubit cumulants are computed from the $1$- and $2$-qubit conditional probabilities:
\begin{equation}
\begin{split}
\lambda_{a,b}(x_j^{(a)}x_j^{(b)} | x_i^{(a)}x_i^{(b)}) &=
p_{a,b}(x_j^{(a)}x_j^{(b)} | x_i^{(a)}x_i^{(b)}) \\
&- p_a(x_j^{(a)} | x_i^{(a)}) p_b(x_j^{(b)} | x_i^{(b)}).  
\end{split}
\label{eq:two_body_term}
\end{equation}

\subsection{Cumulant construction of MFMs}
\label{sec:cumulant_method}
We assume that $1$- and $2$-qubit terms are the dominant terms when constructing an $n$-qubit MFM $K_{q_0, ..., \{q_{n-1}\}}$, thus higher order terms in the cumulant expansion are constructed using the terms in Eqs. \ref{eq:one_body_terms} and \ref{eq:two_body_term}.
We show the third-order cumulant term as an example.  Starting from the $3$-qubit conditional probabilities: 
\begin{equation}
\begin{split}
\lambda_{abc}(x_j^{(a)}x_j^{(b)}x_j^{(c)}& | x_i^{(a)}x_i^{(b)}x_i^{(c)}) = p_{abc}(x_j^{(a)}x_j^{(b)}x_j^{(c)} | x_i^{(a)}x_i^{(b)}x_i^{(c)})\\
&- \biggl[ p_{a}(x_j^{(a)} | x_i^{(a)}) p_{bc}(x_j^{(b)}x_j^{(c)} | (x_i^{(b)}x_i^{(c)}) \\
&+ p_{c}(x_j^{(c)} | x_i^{(c)}) p_{ab}(x_j^{(a)}x_j^{(b)} | (x_i^{(a)}x_i^{(b)}) \\
&+  p_{b}(x_j^{(b)} | x_i^{(b)}) p_{ca}(x_j^{(c)}x_j^{(a)} | (x_i^{(c)}x_i^{(a)}) \biggr]  \\
&+ 2 p_{a}(x_j^{(a)} | x_i^{(a)})p_{b}(x_j^{(b)} | x_i^{(b)})p_{c}(x_j^{(c)} | x_i^{(c)}),
\end{split}
\label{eq:third_order_cumulant}
\end{equation} 
then setting the left-hand side of Eq. \ref{eq:third_order_cumulant} to zero we can compute element-by-element the conditional probabilities $\Tilde{p}$ of a composite MFM $\Tilde{K}_{abc}$ in terms of the $1$- and $2$-qubit cumulants in Eqs \ref{eq:one_body_terms} and \ref{eq:two_body_term}.

\subsection{Extracting MFMs}
\label{sec:spectators_and_extracted}
The $1$- and $2$-qubit MFMs measured on hardware are the basic elements used in Eq. \ref{eq:two_body_term}.  However this approach can fail to capture higher-order correlated noise (see Section \ref{sec:discussion}).  Throughout this work we utilize two additional methods for generating noisy estimates of $2$-qubit MFMs.  The first method exploits classical statistics, the second method using additional hardware qubits.

The MFM is a matrix of classical conditional probabilities $p(x_j|x_i)$ can be marginalized over in order to estimate the conditional probabilities on a smaller $m$-qubit sub-system. In other words, if the full MFM measured over $n$-qubits is known, we can extract an estimated MFM on $m<n$ qubits by marginalizing over the rows and columns, for example:
\begin{equation}
p_{ac} = \sum_{x_i^{b},x_j^{b}}(x_j^{(a)}x_j^{(b)}x_j^{(c)} | x_i^{(a)}x_i^{(b)}x_i^{(c)}).
\end{equation}
This method is used in Section \ref{sec:discussion} to demonstrate the cumulant method for constructing the full MFM using estimated sub-system MFMs.

A second approach to estimating subsystem MFMs is an adaptation of the spectator qubit method introduced in \cite{sun2018efficient,geller2020efficient}. as follows: the ($\binom{n}{2}$) $2$-qubit terms for a $n$-qubit layout are measured with Hadamard gates executed on the remaining $(n-2)$ qubits.  Then from the full measured distribution over $2^n$ states we extract $2$-qubit MFMs.  The addition of spectator qubits does not increase the number of circuits needed to evaluate each $2$-qubit MFM and the cost for measuring all $2$-qubit terms is (4$\binom{n}{2}$).  

\subsection{MFM cluster products}
\label{sec:cluster_method}
The $2$-body cumulant in Eq. \ref{eq:two_body_term} could be rewritten using MFMs of qubit \emph{clusters} as the base factors.  For example, the MFM for a compound system $AB$ may be approximated as $K_A \otimes K_B$.  In this paper we define clusters as subsets of qubits that are connected on the hardware graph, but the connectivity is not a strict requirement. 

This approach captures correlations of all orders within each cluster, but ignores correlations between clusters.  Such a partitioning of qubits into clusters can often be often motivated by physical principles. The cumulant for the full MFM (i.e., the difference between the full MFM and the product of cluster MFMs) then serves as a metric for the degree of correlation between clusters of hardware qubits. Such a metric may be used to decide which clusters are least correlated and thus most suitable for multiplexing.   In Section \ref{sec:cluster_MFM} we investigate $n$-qubit MFMs constructed from clusters of size $\approx n/2$.

\subsection{Metrics}
We quantify the accuracy of a full MFM reconstruction using $3$ measures which are computed between matrix elements (\metricId{}, \metricK{}) or between the diagonal matrix elements (\metricF{}).

The MFM reduces to the identity matrix in the absence of errors.  The total amount of hardware error in a MFM has been quantified previously by $\Vert (K - \mathbbm{1}) \Vert_F$, where $\Vert \cdot \Vert_F$ denotes the Frobenius matrix norm \cite{hamilton2019error}.  To assist in comparison between MFMs of different size, we note that \metricId{} is related to the RMSE of ($K -\mathbbm{1})$ by a factor of $1/\sqrt{2^n}$.
We also define \metricK{} to quantify the closeness of $K$ and $\Tilde{K}$.  For some purposes the type of error may not be of interest and only the fidelity $f_i$ of each target state $x_i$ is of interest. (Our fidelity is equivalent to the probability of successful trial (PST) introduced in \cite{tannu2019mitigating}.) For this we use \metricF, the RMSE of the fidelity differences, 
\begin{equation}
\Delta f = \sqrt{\frac{1}{2^n} \sum_{i=1}^{2^n} (f_i^{\tilde{K}} - f_i^{K})^2}
\end{equation}

The scalar correlation factor (SCF) is a separate metric which quantifies the degree of independence between two qubits or clusters ($A,B$) using the Frobenius norm of the two-body cumulant matrix introduced in Eq. \ref{eq:two_body_term}:
\begin{equation}
 \Lambda_{AB} \equiv \Vert \lambda_{AB}\Vert_F = \sqrt{ \sum_{ij} |\lambda_{AB}(i,j)|^2}.
\label{eq:SCF}
\end{equation}
where the sum is over distinct input and output states of composite system $AB$. $\Vert \lambda_{AB}\Vert_F$ may be understood as (proportional to) the root-mean-square error of conditional probabilities when $K_{AB}$ is approximated by $K_{A} \otimes K_{B}$. Theorem 1 of \cite{kubocumulant} and its corollary state that a cumulant $\lambda_{AB}(j|i)$ is zero if and only if the variables of $A,B$ can be divided into two or more statistically independent groups.  Thus $\Vert \lambda_{AB}\Vert_F$ may be interpreted as the degree of statistical correlation between subsystems $A,B$.  We will use this metric to quantify correlations between qubit clusters in Section \ref{sec:cluster_MFM} and individual qubits in Section \ref{sec:discussion}.

\subsection{Uncertainty Analysis}
\label{sec:uncertainty_prop}
The main source of uncertainty is the randomness of quantum measurement outcomes.  Each row of a MFM is generated by preparing a specific quantum state - the true output distribution $p$ is obtained by preparing and sampling the state many times and estimating $p(x_j|x_i)$ as the fraction of samples that were $x_j$ when the state $x_i$ was prepared.
If $n_s$ samples are taken, the uncertainty in $p(x_j|x_i)$ is
\begin{equation}
    \sigma (p(x_j|x_i)) = \frac{1}{n_s} p(x_j|x_i)(1-p(x_j|x_i)).
\end{equation}
Uncertainty in the elements of $K$ translates into uncertainty in the elements of $\lambda_{ab}$. To distinguish correlations in the conditional probabilities constructed via the two-body cumulant (\ref{eq:two_body_term}) from statistical noise we use standard propagation of uncertainty to establish a lower bound on the terms in the SCF.

For instance, if the MFMs $K_a$, $K_b$, and $K_{ab}$ for a pair of qubits $a,b$ are estimated from independent measurements, it can be shown that the uncertainty in $\lambda_{ab}(j|i)$ is (to lowest order in statistical fluctuations) upper bounded by
\begin{equation}
    \sigma(\lambda_{ab}(j|i)) \le \left( \frac{p_{ab}(x_j|x_i) + p_a(x_j|x_i) + p_b(x_j|x_i)}{n_s} \right)^{1/2}.
\end{equation}
The corresponding uncertainty in $\Lambda_{ab}$ (eq.\ \ref{eq:SCF}) is upper bounded (to lowest order in statistical fluctuations) by 
\begin{equation}
    \sigma(\Lambda_{ab}) \leq \left( \frac{1}{\Lambda_{ab}^2} \sum_{i,j} \lambda_{ab}(j|i)^2 \sigma(\lambda_{ab}(j|i))^2 \right)^{1/2}.
\label{eq:sigma_Lamb}
\end{equation}
We consider the correlation between a pair of qubits $(a,b)$ to be statistically significant if the entries of the SCF satisfy: $\Lambda_{ab} > \sigma(\Lambda_{ab})$.

If $K_a$ and $K_b$ are extracted from a larger MFM (using either method introduced in Section \ref{sec:spectators_and_extracted}) instead of being measured in separate experiments then the cumulant uncertainties in this case are approximately the same as in the previous case and satisfy the bounds above.  However, in this case each cumulant $\lambda_{ab}(j|i)$ must be scaled by $(1-1/n_s)^{-1}$ to account for statistical correlations between the joint and marginal distributions and obtain an unbiased estimate.

As the number of qubits grows, it is intuitive that a higher shot size will be needed in order to ensure the final state is sufficiently sampled.  In order to assess the relative error introduced in reconstructed and measured MFMs for a given shot size, we sampled single-qubit MFMs, and $n$-qubit MFMs from a simulated quasi-ideal chip consisting of independent, identical qubits with the same $p(x_j|x_i)$.  This allows for the comparison of $\tilde{K}$ and $K$ to the ideal MFM, $K_{I}$ obtained from $p(x_j|x_i)$.  For collections of up to 20 qubits, with $.9<p(0|0) = p(1|1)<.98$, and fixed shot size $n_s$=8192, we performed this sampling until a stable distribution of $\Vert (K - K_I) \Vert_F$, and $\Vert (\tilde{K} - K_I) \Vert_F$ are obtained and then used to gauge the relative uncertainty.  For all case sizes explored in this work, this analysis indicated that for a given shot size, reconstructed MFMs perform better than measured, and that the benefit increases with qubit number.  While this was only for single-qubit reconstructions, we expect that a similar analysis of more complicated reconstructions would yield similar findings.

\subsection{Computational Cost}
\begin{figure}[htbp]
  \centering
  \includegraphics[width=\columnwidth]{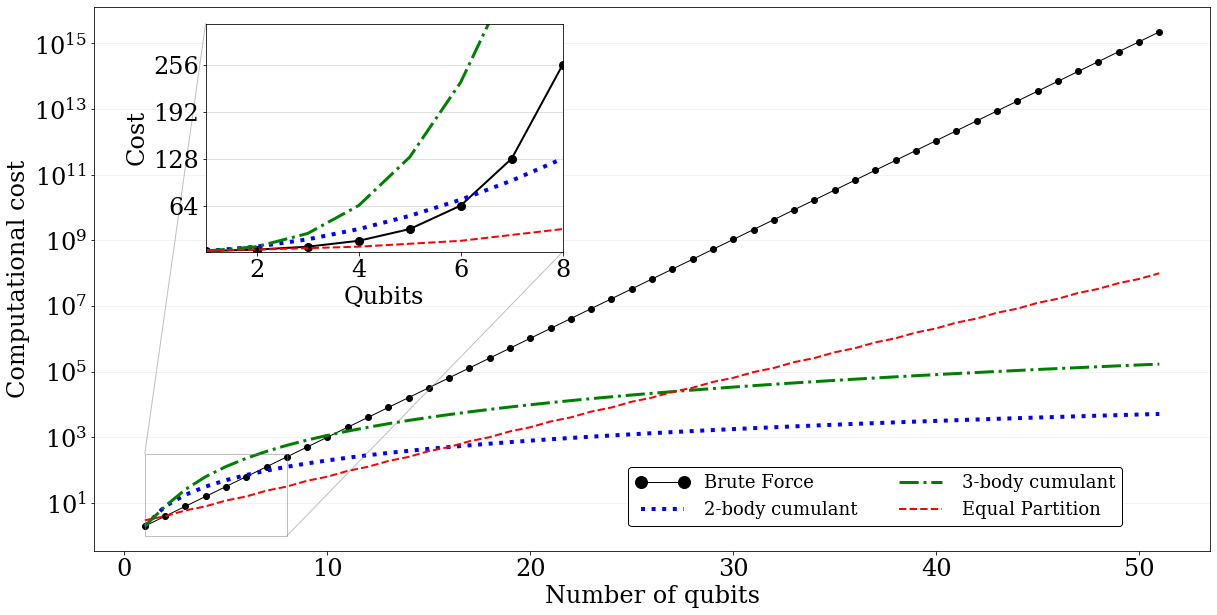}
  \caption{Computational cost to construct a $n$-qubit MFM using full measurements (black, circles), a $2$-body cumulant expansion (blue), a $3$-body cumulant expansion (green) or bipartition MFMs (red, dashed) plotted on a logarithmic scale.  (Inset) Detail of scaling for up to $8$ qubits plotted on a linear scale.}
  \label{fig:scaling_cost}
\end{figure}

We define computational cost as the number of circuits that need to be executed on the hardware.  Each single qubit MFM requires $2$ circuits to be executed, each $2$-qubit MFM requires $4$ circuits and the full construction of a $n$-qubit MFM requires $2^n$ circuits.  
Fig. \ref{fig:scaling_cost} shows how the computational cost required to construct an $n$-qubit MFM scales for different construction methods.  Assuming a fixed number of samples ($n_s$), for each method the number of circuits required are: $2^n$ for measurement of the full MFM; $4\binom{n}{2}$ for construction using cumulants measured up to order 2; and $8\binom{n}{3})$ for construction using cumulants measured up to order 3.  We also include the scaling of a non-cumulant method in which a MFM is constructed from two (approximately) equal sub-regions of size $k$ and $n-k$, with cost $2^{k} + 2^{n-k}$. 

Even though adding higher order terms into the cumulant expansion incurs additional cost, once the number of qubits in the MFM exceeds $8$ qubits both cumulant expansion methods and the regional method have a lower computational overhead than the exact MFM construction (see Fig.~\ref{fig:scaling_cost}). However for MFMs on $n < 5$ qubits the exact construction is less expensive that the cumulant methods.

To measure a full $5$-qubit MFM the circuit cost is $2^5 = 32$.  Measuring all single qubit MFMs ($q_{a}$) requires $10$ circuits, while a construction based on measurement of all $2$-qubit subsystems has a cost of $40$ circuits.  The cost of a MFM constructed from a $2$-qubit MFM and $3$-qubit MFM is $12$, while the cost of a MFM constructed from a $1$-qubit MFM and $4$-qubit MFM is $18$.

\subsection{Hardware layouts}
The data and results presented in this paper cover the construction of MFMs on up to $8$ qubits using shallow circuits.  Results on $5$-qubit MFMs were constructed in Qiskit.  Results on $6$ and $8$-qubit MFMs were measured using an XACC implementation \cite{mccaskey2020xacc} (for details see Appendix \ref{appendix:xacc_implementation}).  MFMs are defined with respect to the computational basis (Z basis) using circuits constructed with only the native $X$ gate implemented via the fixed rotation gate $\mathtt{u3}(\pi,0,\pi)$. In Section \ref{sec:cumulant_MFM} we report data measured with spectator qubits (see Section \ref{sec:spectators_and_extracted}) in which Hadamard gates are executed on the unmeasured qubits ("spectator qubits") to randomize any potential influence they might have in the extracted MFMs.

\begin{figure}[htbp]
  \centering
  \includegraphics[width=0.75\columnwidth]{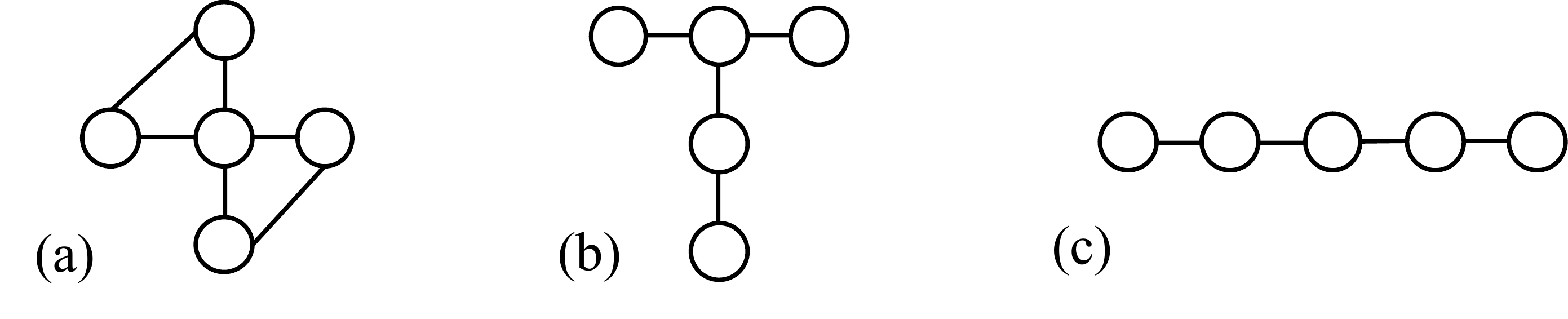}
  \caption{Qubit layouts for the $5$ qubit MFMs: (a) bowtie (\YORK), (b) tree (\VAL{} or subsets ($T_1,T2$) on \CAM{}), (c) simple chain (subset $C_0$ on \CAM).}
  \label{fig:5qubit_backends}
\end{figure}

\begin{figure}[htbp]
  \centering
  \includegraphics[width=0.9\columnwidth]{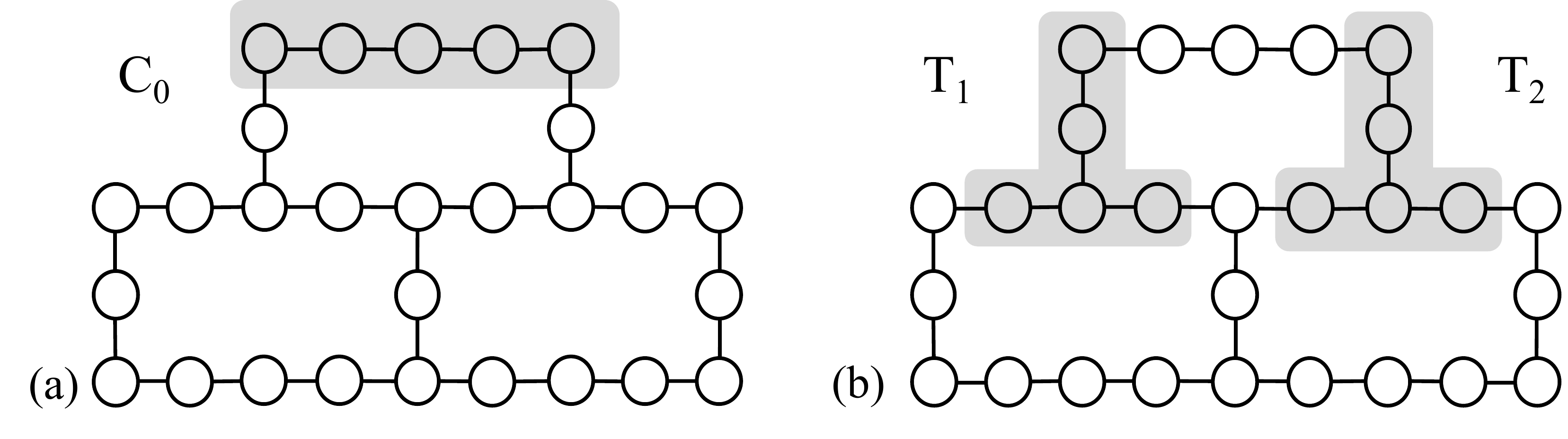}
  \caption{Subsets used to measure $5$-qubit kernels on \CAM{} (grey, shaded): (a) a $5$-qubit chain ($C_0$) (b) two $5$-qubit trees ($T_1$,$T_2$)}
  \label{fig:cambridge_5qubit_subsets}
\end{figure}

The MFM results presented in Secs.~\ref{sec:cumulant_MFM} and \ref{sec:cluster_MFM} were generated on multiple superconducting qubit devices (QPUs)  available from IBM.  Most devices were accessed via cloud-based priority-queues, except for \VAL{} which was accessed via a dedicated queue. In this paper each QPU is represented by a graph that schematically illustrates the physical layout of the qubits and controllable couplings between them. Using the $5$-qubit devices \VAL{}, and \YORK{} we measured $5$-qubit MFMs that cover the entire QPU (see Fig. \ref{fig:5qubit_backends}).  On the $28$ qubit device \CAM{} we measured MFMs for selected $5$-qubit subsets (see Fig. \ref{fig:cambridge_5qubit_subsets}).  On the $20$-qubit devices \BOBO{} and \JOJO{} we measured $6$- and $8$-qubit MFMs.  

\section{Cumulant series construction of MFM}
\label{sec:cumulant_MFM}
In this section we reconstruct of $5$ and $6$ qubit MFMs using measurements on $1$ and $2$ qubits.  We report the accuracy of our reconstructions using \metricId{} and \metricF{} values.

Our results were obtained using circuits executed on the $5$-qubit devices \VAL{} and \YORK{}, the $28$-qubit device \CAM{} and the $20$-qubit device \JOJO.  On \CAM{} we used three separate qubit subsets, shown in Fig. \ref{fig:cambridge_5qubit_subsets}.  On \JOJO{} we used an $8$-qubit layout ($J_8$) shown in Fig. \ref{fig:20qubit_backends}.  

\begin{figure}[htbp]
  \centering
  \includegraphics[width=0.9\columnwidth]{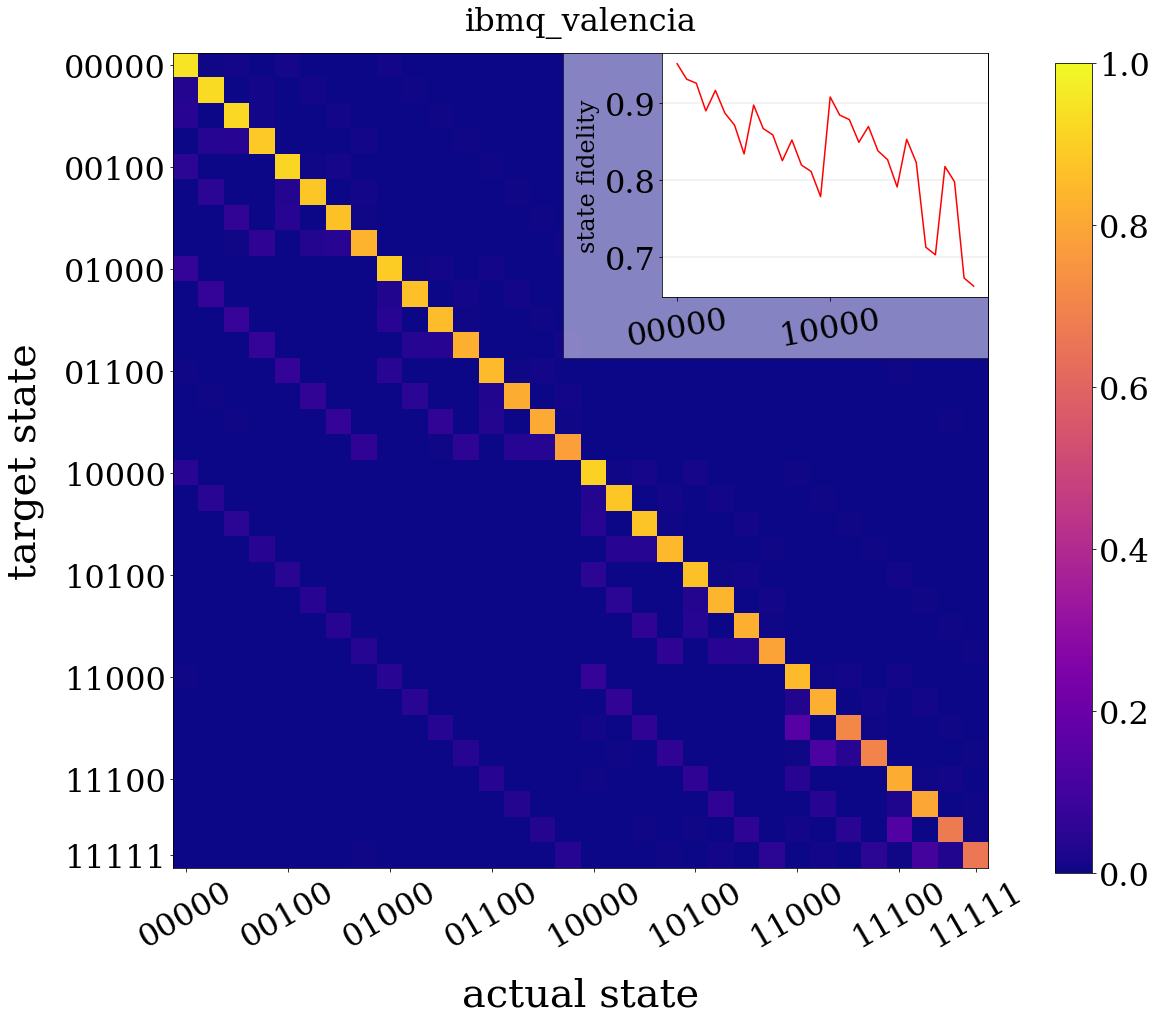}
  \caption{Full $5$-qubit MFM measured on \VAL{} (March 13, 2020).  This MFM returned one of the lowest values of \metricId.  The inset plots highlight the individual state fidelities which are also plotted along the kernel diagonal. }
  \label{fig:valencia_5qubit_measured}
\end{figure}

For each target $n$-qubit MFM we measured all 1-qubit MFMs and all $\binom{n}{2}$ $2$-qubit MFMs. We also measured the full $5$-qubit MFM of each device for comparison.  An example of a full measured MFM, measured on \VAL{} is shown in Fig.~\ref{fig:valencia_5qubit_measured}.  

As an initial test we constructed approximate MFMs using only vendor-provided single-qubit calibration data.  For each qubit the conditional probabilities $p(1 | 0)$ (the probability of measuring a qubit in state $|1\rangle$ when prepared in state $|0\rangle$) and $p(0 | 1)$ (the probability of measuring a qubit in state $|0\rangle$ when prepared in state $|1\rangle$) can be retrieved using the \texttt{properties} methods associated with the backends (access provided by IBM Quantum through the IBM Quantum Provider).  From these two values one can construct the $1$-qubit MFM
\begin{equation}
K^{\mathrm{VC}} = 
\begin{pmatrix}
1-p(1 | 0) & p(1 | 0) \\
p(0 | 1) & 1-p(0 | 1).
\end{pmatrix}
\label{mat:basic_1q_kernel}
\end{equation}
(We use the superscript ${}^\mathrm{VC}$ to denote an MFM constructed from vendor calibration data, as opposed to one directly measured on hardware.) In our experiments, calibration information was collected and stored at the start of each set of measurements of the full $5$-qubit MFM.  The 5-qubit MFM was then approximated as $K_0^{\mathrm{VC}} \otimes \cdots \otimes K_4^{\mathrm{VC}}$.
Albeit simplistic, constructing an MFM from calibration data does not require the end user to execute any circuits on a backend. 

We also constructed approximate MFMs using the second-order cumulant expansion (Section \ref{sec:cumulant_exp}) and measurement of all $1$- and $2$-qubit MFMs. The kernel constructed in this way is denoted $K_{\binom{Q}{1} + \binom{Q}{2}}$, where $Q=\{0,1,2,3,4\}$ denotes the full set of qubits.  The \metricId{}, \metricK{} and \metricF{} values for each type of approximate MFM generated for each device are presented in Tables \ref{tab:TABLE1_cumulant_nospec_5qubit_chip} and \ref{tab:TABLE2_5qubit_cumulant_subsets}.
{\renewcommand{\arraystretch}{1.5}%
 \begin{table}
      \centering
      \caption{Accuracy of 5-qubit MFMs constructed from 1- and 2-qubit MFMs directly measured without spectator qubits.}
     \label{tab:TABLE1_cumulant_nospec_5qubit_chip}
     \begin{tabular}{|p{2.8cm}|p{1.5cm}|p{1.7cm}|p{1.1cm}|}
     \hline
     \multicolumn{4}{|c|}{ \VAL{} } \\ \hline
\tabhead
$K \equiv \qkern{0,1,2,3,4}$ & 1.13 & 0.0 & 0.0  \\ \hline
$K^{\mathrm{VC}}_0 \otimes \cdots \otimes K^{\mathrm{VC}}_4$ & 0.658 & 0.560 & 0.084 \\ \hline
$K_0 \otimes \cdots \otimes K_4$ & 0.980 & 0.285 & 0.036  \\ \hline
$K_{\binom{Q}{1} + \binom{Q}{2}}$ & 0.997  & 0.279 & 0.036  \\ \hline
\hline
\multicolumn{4}{|c|}{ \YORK{} } \\ \hline
\tabhead
$K \equiv \qkern{0,1,2,3,4}$ & 1.296 & 0.0 & 0.0 \\ \hline
$K^{\mathrm{VC}}_0 \otimes \cdots \otimes K^{\mathrm{VC}}_4$ & 0.634 & 0.897 & 0.122  \\ \hline
$K_0 \otimes \cdots \otimes K_4$ & 0.550 & 0.969 & 0.136  \\ \hline
$K_{\binom{Q}{1} + \binom{Q}{2}}$  & 0.615 & 0.926 & 0.126  \\ \hline
     \end{tabular}
 \end{table}
}
{\renewcommand{\arraystretch}{1.5}%
 \begin{table}
      \centering
      \caption{Accuracy of $5$ and $8$-qubit MFMs constructed from $1$- and $2$-qubit MFMs}
     \label{tab:TABLE2_5qubit_cumulant_subsets}
     \begin{tabular}{|p{2.8cm}|p{1.5cm}|p{1.7cm}|p{1.1cm}|}
     \hline
     \multicolumn{4}{|c|}{ \CAM{}, region $T_1 = \{8,9,10,5,0\}$ } \\ \hline
\tabhead
$K \equiv K_{T_1}$ & 3.616 & 0.0 & 0.0 \\ \hline
$\bigotimes_{q\in T_1} K^{\mathrm{VC}}_{q}$ & 2.650  & 1.367 & 0.201 \\ \hline
$\bigotimes_{q\in T_1} K_{q}$ & 2.917 & 0.965 & 0.138 \\ \hline
$K_{\binom{T_1}{1} + \binom{T_1}{2}}$ & 2.817 & 1.219 & 0.138 \\ \hline
\hline
     \multicolumn{4}{|c|}{ \CAM{}, region $T_2 = \{12,13,14,6,4\}$ } \\ \hline
\tabhead
$K \equiv K_{T_2}$ & 1.781 & 0.0 & 0.0\\ \hline
$\bigotimes_{q\in T_2} K^{\mathrm{VC}}_{q}$ & 1.371 & 0.504 & 0.072  \\ \hline
$\bigotimes_{q\in T_2} K_{q}$ & 1.541 & 0.409 & 0.051 \\ \hline
$K_{\binom{T_2}{1} + \binom{T_2}{2}}$ & 1.653 & 0.284 & 0.035 \\ \hline
\hline
     \multicolumn{4}{|c|}{ \CAM{}, region $C_0 = \{0,1,2,3,4\}$ } \\ \hline
\tabhead
$K \equiv K_{C_0}$ & 2.247 & 0.0 & 0.0\\ \hline
$\bigotimes_{q\in C_0} K^{\mathrm{VC}}_{q}$ & 2.129  & 0.137 & 0.019 \\ \hline
$\bigotimes_{q\in C_0} K_{q}$ & 2.263  & 0.173 & 0.017 \\ \hline
$K_{\binom{C_0}{1} + \binom{C_0}{2}}$ & 2.149 & 0.623 & 0.015 \\ \hline
\hline
\multicolumn{4}{|c|}{ \JOJO{}, qubits $J_8=\{6,5,10,11,8,9,13,14\}$ } \\ \hline
\tabhead
$\qkern{6,5,10,11,8,9,13,14}$ &  3.889 & 0.0 & 0.0\\ \hline
$\bigotimes_{q\in J_8} K^{\mathrm{VC}}_{q}$ & 4.639  & 1.382 & 0.056 \\ \hline
$\bigotimes_{q\in J_8} K_{q}$ & 3.649 & 0.536 & 0.003 \\ \hline
$K_{\binom{J_8}{1} + \binom{J_8}{2}}$ & 5.183 & 3.0293 & 0.002 \\ \hline
     \end{tabular}
 \end{table}
 }
\subsection{Subsystems measured with spectator qubits}
In this section we present results on an alternate application using the cumulant expansion method, in which have additional noise in the $1$ and$2$-qubit subsystem MFMs before constructing the $5$, $6$ and $8$-qubit MFMs.  We adapt the spectator qubit method introduced in \cite{sun2018efficient,geller2020efficient} as follows: the ($n \choose 2$) $2$-qubit terms for a $n$-qubit layout are measured with the addition of Hadamard gates on the remaining $(n-2)$ qubits.  From the full measured distribution over $2^n$ states we extract $2$-qubit MFMs.  The addition of spectator qubits does not increase the number of circuits needed to evaluate each $2$-qubit MFM and the cost for measuring all $2$-qubit terms is (4$n \choose 2$).

{\renewcommand{\arraystretch}{1.5}%
 \begin{table}
      \centering
      \caption{Accuracy of 5-qubit and 8-qubit MFMs constructed from 1- and 2-qubit MFMs measured with spectator qubits.}
     \label{tab:TABLE3_cumulant_spec_5qubit_8qubit}
     \begin{tabular}{|p{2.8cm}|p{1.5cm}|p{1.7cm}|p{1.1cm}|}
     \hline
     \multicolumn{4}{|c|}{ \VAL{} } \\ \hline
\tabhead
$K \equiv \qkern{0,1,2,3,4}$ & 1.12 & 0.0 & 0.0\\ \hline
$\bigotimes K^{\mathrm{VC}}_{q}$ & 1.17 & 0.347 & 0.043  \\ \hline
$K_0 \otimes \cdots \otimes K_4$ & 1.04 & 0.260 & 0.034\\ \hline
$K_{\binom{Q}{1} + \binom{Q}{2}}$ & 1.08 & 0.976 & 0.030\\ \hline
\hline
\multicolumn{4}{|c|}{ \YORK{}} \\ \hline
\tabhead
$K \equiv \qkern{0,1,2,3,4}$ & 1.65 & 0.0 & 0.0\\ \hline
$\bigotimes K^{\mathrm{VC}}_{q}$ & 0.674 & 1.105 & 0.168  \\ \hline
$K_0 \otimes \cdots \otimes K_4$ & 1.622 & 0.734 & 0.060 \\ \hline
$K_{\binom{Q}{1} + \binom{Q}{2}}$  & 1.512 & 0.663 & 0.071\\ \hline
\hline
\multicolumn{4}{|c|}{ \CAM{} ($C_0$) } \\ \hline
\tabhead
$K \equiv \qkern{0,1,2,3,4}$ & 2.06 & 0.0 & 0.0\\ \hline
$\bigotimes_{q\in C_0} K^{\mathrm{VC}}_{q}$ & 2.30 & 0.322 & 0.045  \\ \hline
$K_0 \otimes \cdots \otimes K_4$ & 1.96 &  0.165& 0.019\\ \hline
$K_{\binom{Q}{1} + \binom{Q}{2}}$  & 1.97 & 0.474 & 0.009\\ \hline
\hline
\multicolumn{4}{|c|}{ \CAM{} ($T_1$) } \\ \hline
\tabhead
$K \equiv \qkern{0,1,2,3,4}$ & 2.879 & 0.0 & 0.0\\ \hline
$\bigotimes_{q\in T_1} K^{\mathrm{VC}}_{q}$ & 2.426 & 0.958 & 0.118 \\ \hline
$K_0 \otimes \cdots \otimes K_4$ & 3.553 & 0.874 & 0.116\\ \hline
$K_{\binom{Q}{1} + \binom{Q}{2}}$  & 3.375 & 0.885 & 0.112\\ \hline
\hline
\multicolumn{4}{|c|}{ \CAM{} ($T_2$) } \\ \hline
\tabhead
$K \equiv \qkern{0,1,2,3,4}$ & 1.66 & 0.0 & 0.0\\ \hline
$\bigotimes_{q\in T_2} K^{\mathrm{VC}}_{q}$ & 1.463 & 0.419 & 0.041   \\ \hline
$K_0 \otimes \cdots \otimes K_4$ & 2.054 & 0.675 & 0.060 \\ \hline
$K_{\binom{Q}{1} + \binom{Q}{2}}$  & 1.612 & 0.328  & 0.029 \\ \hline
\hline
\multicolumn{4}{|c|}{ \JOJO{}, qubits $J_8=\{6,5,10,11,8,9,13,14\}$  } \\ \hline
\tabhead
$\qkern{6,5,10,11,8,9,13,14}$ &  3.889 & 0.0 & 0.0\\ \hline
$\bigotimes_{q\in J_8} K^{\mathrm{VC}}_{q}$ & 4.639  & 1.382 & 0.056 \\ \hline
$\bigotimes_{q\in J_8} K_{q}$ & 4.697 & 1.0838 & 0.003 \\ \hline
$K_{\binom{J_8}{1} + \binom{J_8}{2}}$ & 3.889 & 4.147 & 0.003 \\ \hline
     \end{tabular}
 \end{table}
}

\section{Cluster product construction of MFM}
\label{sec:cluster_MFM}
In this section we reconstruct $5$, $6$ and $8$ qubit MFMs using the cluster product method described in Sec. \ref{sec:cluster_method}.  We report the accuracy of our reconstructions using the \metricId{} metric and the SCF between clusters (see Eq. \ref{eq:SCF}).

\begin{figure}[htbp]
  \centering
  \includegraphics[width=1.\columnwidth]{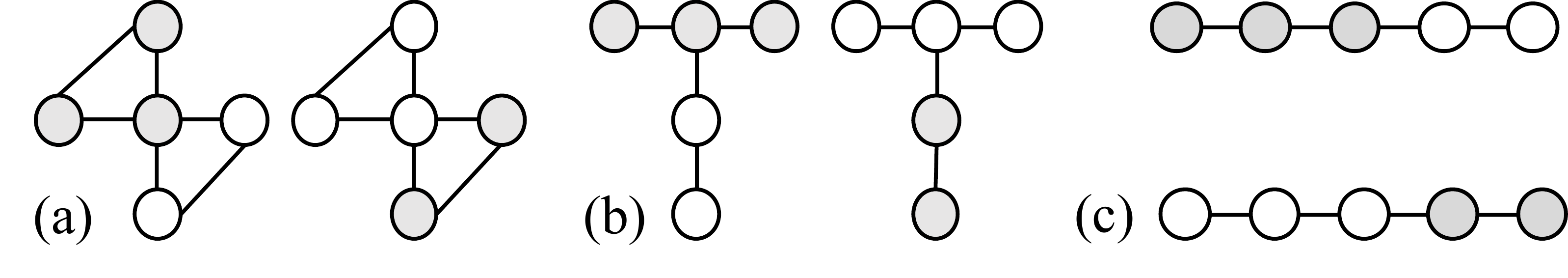}
  \caption{Examples of (3,2) clusters: (a) bowtie (b) tree (c) simple chain.}
  \label{fig:regional_constructions_5qubit}
\end{figure}

\begin{figure}[htbp]
  \centering
  \includegraphics[width=\columnwidth]{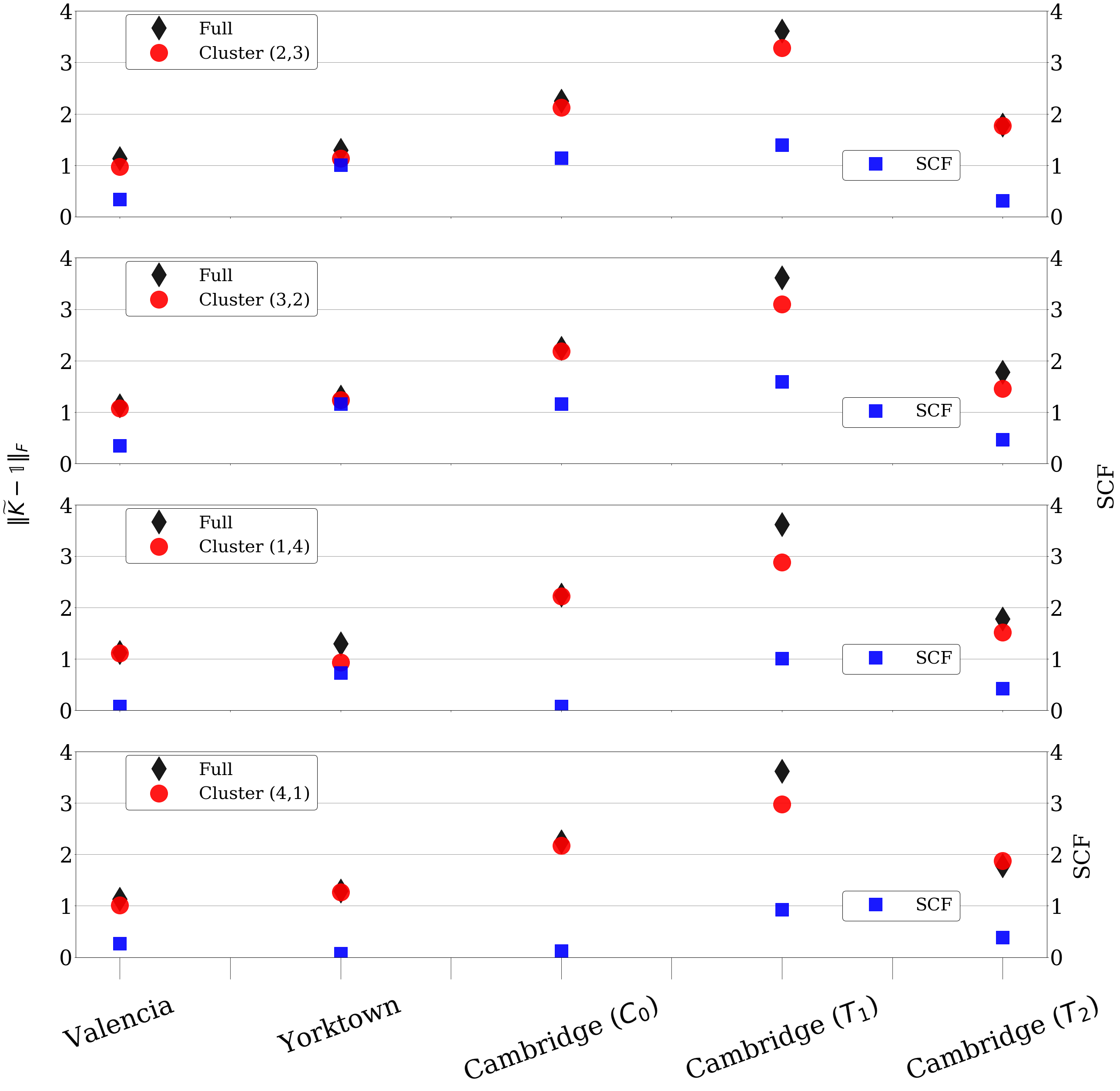}
  \caption{\metricId{} measured on \VAL{}, \YORK{} and \CAM{} using the full $5$-qubit MFM (black, diamonds) compared to MFMs constructed using cluster products (red, circles).  SCF (blue, squares) defined by $\Vert \lambda_{AB} \Vert_F$ between the clusters plotted on secondary axis.}
  \label{fig:metricId_5q_cluster}
\end{figure}

Figure \ref{fig:regional_constructions_5qubit} shows examples of (3,2) qubit clusters for each of the $5$ qubit layouts: tree, bowtie and a simple chain.  We construct our cluster sets such that they respect the bit ordering of the $5$-qubit MFM.  The metrics for the MFMs constructed from these clusters are plotted in Fig. \ref{fig:metricId_5q_cluster}.

\begin{figure}
  \centering
  \includegraphics[width=\columnwidth]{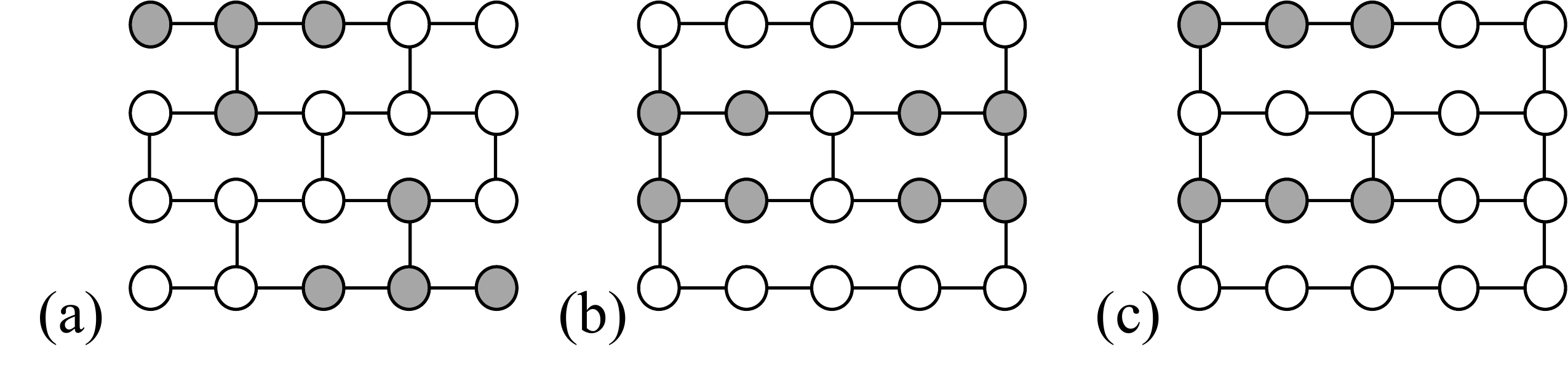}
  \caption{Examples of qubit clusters on $20$-qubit connectivity graphs: (a) $4$-qubit trees shown on \BOBO. (b) $4$-qubit chains ($J_8$) shown on \JOJO (c) $3$-qubit chains shown in \JOJO.}
  \label{fig:20qubit_backends}
\end{figure}

On \BOBO{} and \JOJO{} we reconstructed $6$ and $8$ qubit MFMs using clusters of $3$ and $4$ qubits.  When choosing qubit clusters on the hardware we fixed the layout to be the same (see Fig. \ref{fig:20qubit_backends}).  While all $3$-qubit clusters are chains, the $4$-qubit clusters can be trees or chains.  

\begin{figure}[htbp]
  \centering
  \includegraphics[width=\columnwidth]{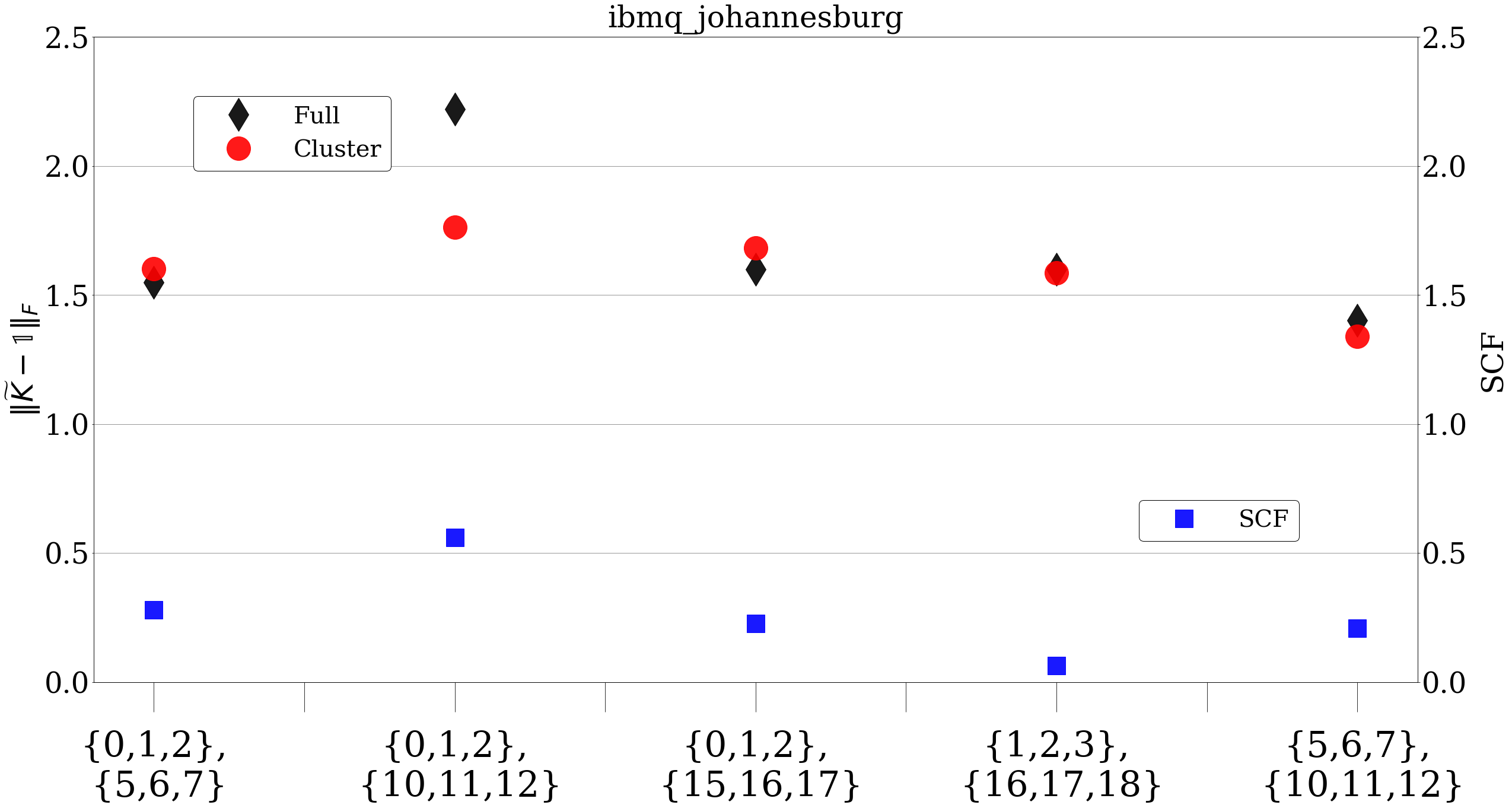}
  \caption{\metricId{} measured on \JOJO{} using the full $6$-qubit MFM (black, diamonds) compared to MFMs constructed using $3$-qubit cluster products (red, circles).  SCF (blue, squares) defined by $\Vert \lambda_{AB} \Vert_F$ between the $3$-qubit clusters used as x-axis labels.}
  \label{fig:metricId_6q_cluster_jojo}
\end{figure}

\begin{figure}[htbp]
  \centering
  \includegraphics[width=\columnwidth]{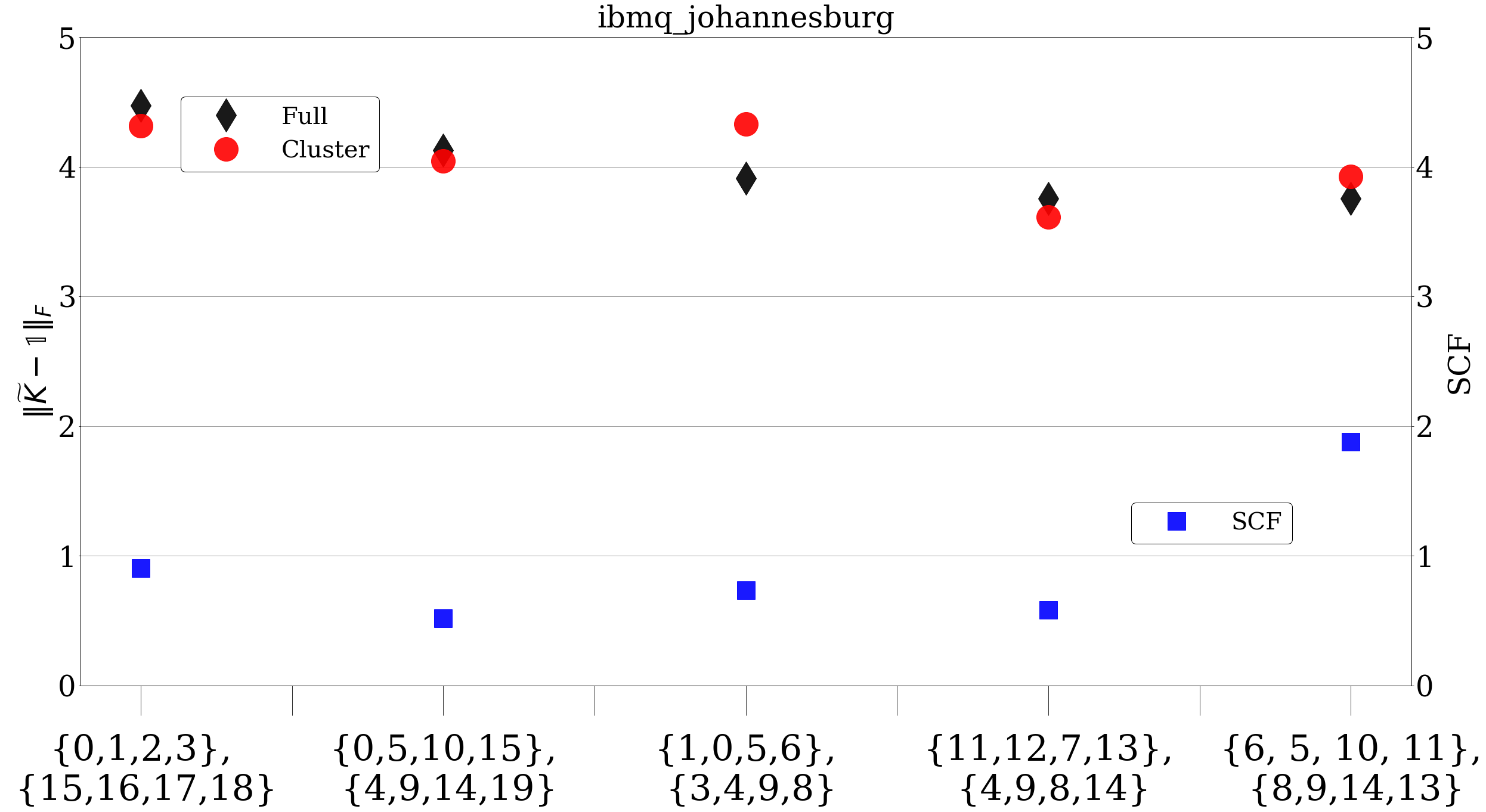}
  \caption{\metricId{} measured on \JOJO{} using the full $8$-qubit MFM (black, diamonds) compared to MFMs constructed using $3$-qubit cluster products (red, circles).  SCF (blue, squares) defined by $\Vert \lambda_{AB} \Vert_F$ between the $4$-qubit clusters used as x-axis labels.}
  \label{fig:metricId_8q_cluster_jojo}
\end{figure}

\begin{figure}
  \centering
  \includegraphics[width=\columnwidth]{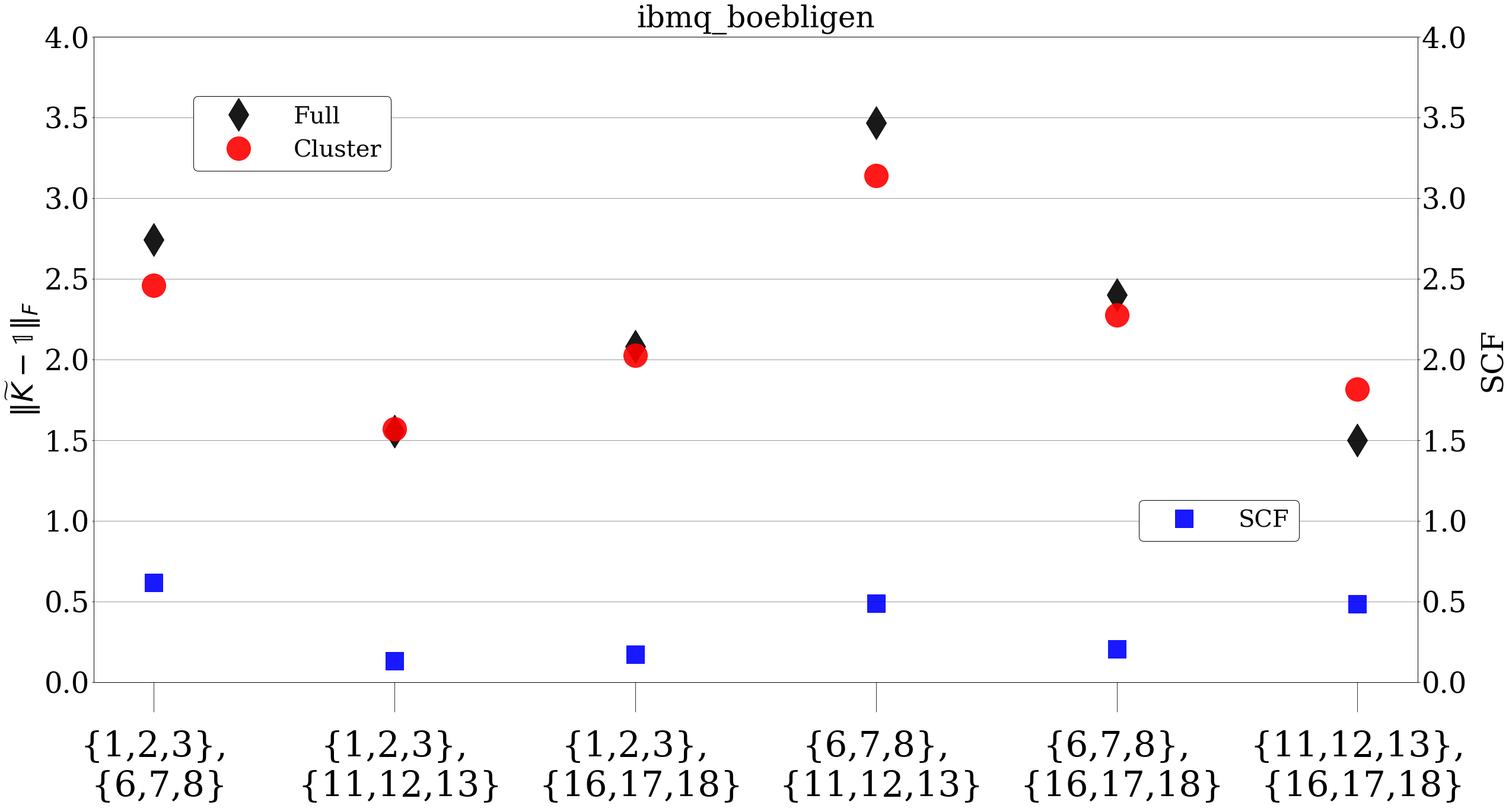}
  \caption{\metricId{} measured on \BOBO{} using the full $6$-qubit MFM (black, diamonds) compared to MFMs constructed using $3$-qubit cluster products (red, circles). SCF (blue, squares) defined by $\Vert \lambda_{AB} \Vert_F$ between the $4$-qubit clusters used as x-axis labels..}
  \label{fig:metricId_6q_cluster_bobo}
\end{figure}

\begin{figure}
  \centering
  \includegraphics[width=\columnwidth]{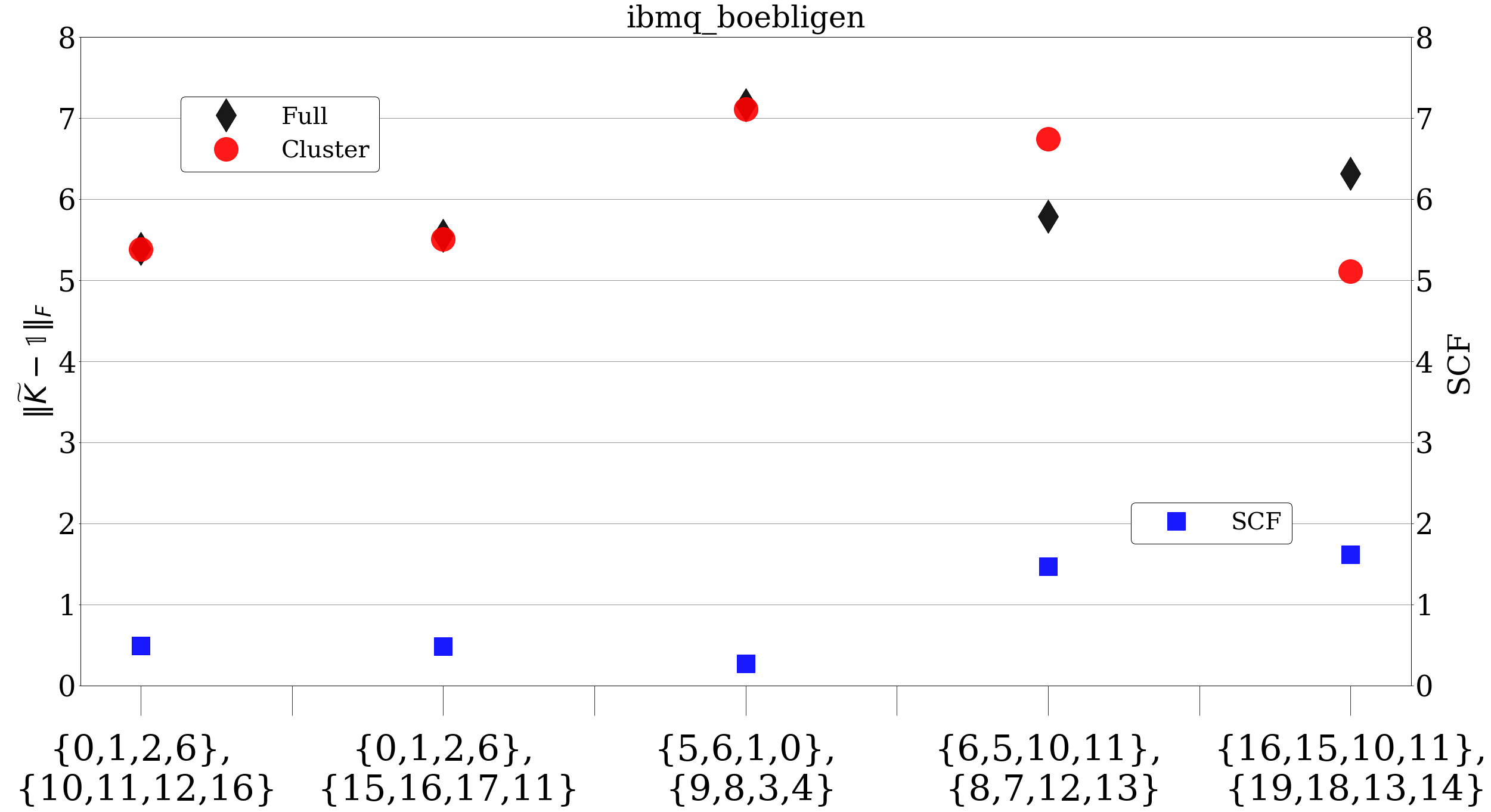}
  \caption{\metricId{} measured on \BOBO{} using the full $8$-qubit MFM (black, diamonds) compared to MFMs constructed using $4$-qubit cluster products (red, circles).  SCF (blue, squares) defined by $\Vert \lambda_{AB} \Vert_F$ between the $4$-qubit clusters used as x-axis labels.}
  \label{fig:metricId_8q_cluster_bobo}
\end{figure}

The data collected on \JOJO{} is plotted in Figs. \ref{fig:metricId_6q_cluster_jojo} and \ref{fig:metricId_8q_cluster_jojo} and the metrics evaluated using data collected on \BOBO{} is plotted in Figs. \ref{fig:metricId_6q_cluster_bobo} and \ref{fig:metricId_8q_cluster_bobo}.

We re-introduce the SCF described in Section \ref{sec:methods} to quantify correlations between regions on a QPU using a $n$ qubit layout decomposed into $2$ disjoint clusters.  Consider a layout of $n$ qubits that is decomposed into two disjoint clusters $A$, $B$ each consisting of approximately $n/2$ qubits.
Eq.\ \ref{eq:two_body_term} may be applied to the two clusters, yielding the cluster cumulant
\begin{equation*}
\begin{split}
    \lambda_{AB}(j|i) &= p_{AB}(x_j^{(A)} x_j^{(B)} | x_i^{(A)} x_i^{(B)}) \\
    &- p_A(x_j^{(A)} | x_i^{(A)}) p_B(x_j^{(B)} | x_i^{(B)})
\end{split}
\end{equation*}
and the corresponding SCF $\Lambda_{AB} = \Vert \lambda_{AB}\Vert_F$.

In Figs. ~\ref{fig:metricId_6q_cluster_jojo},\ref{fig:metricId_8q_cluster_jojo},\ref{fig:metricId_6q_cluster_bobo},\ref{fig:metricId_8q_cluster_bobo} we plot the SCF value $\Lambda_{AB}$ on the second y-axis. The full MFM is measured with a specific ordering of hardware qubits, which must be respected in the ordering of the clusters.  In order to preserve the proper qubit ordering the rows and columns of the full MFM may have to be transposed.  For example, using the layout $\{ 0,1,2,15,16,17 \}$ from Fig \ref{fig:metricId_6q_cluster_jojo},  $\Lambda_{AB} = \Vert K_{\{0,1,2,15,16,17\}} - K_{\{0,1,2\}}\otimes K_{\{15,16,17\}} \Vert_F$. However to evaluate $\Lambda_{AB}$ the rows and columns of $K_{\{0,1,2,15,16,17\}}$ must be transposed to yield $K_{\{15,16,17,0,1,2\}}$. 

\section{Discussion}
\label{sec:discussion}
In Sections \ref{sec:cumulant_MFM} and \ref{sec:cluster_MFM} we constructed $n$-qubit MFMs using tensor products of single-qubit MFMs, a cumulant-based method, and also with cluster MFMs.  The accuracy of the constructed MFMs was quantified by \metricId.  There is a general upper bound to this value, introduced previously in \cite{hamilton2019error}: if $\Tilde{K}$ is purely white noise (each row is a uniform distribution over all $2^n$ states) then $\Vert \Delta_1 \Tilde{K} \Vert_F = \sqrt{(2^n-1)}$.  

\subsection{Direct measurement of sub-system MFMs}
Overall, the accuracy of the MFM construction method is not dependent on the total noise in the full MFM.  Rather the accuracy of the reconstruction method is dependent on the degree of correlations that are captured in the sub-system MFMs.  If a sub-system MFM of $m$ qubits is measured without the addition of spectator qubits, then the highest degree of correlations contained in that MFM is order $m$.  Reconstructed MFMs using single qubit measurements were the least accurate - $K_q^{VC}$ or $K_q$ underestimated the general noise of the full MFM (see Tables \ref{tab:TABLE1_cumulant_nospec_5qubit_chip} and \ref{tab:TABLE2_5qubit_cumulant_subsets}).  

The addition of $2$-qubit MFMs did not lead to conclusive improvement in accuracy- for \CAM{} ($C_0, T_1$) and \JOJO ($J_8$) the construction using $K_{\binom{Q}{1} + \binom{Q}{2}}$ resulted in a MFM with a much lower value of \metricId. Likewise adding spectator qubits in the $1$ and $2$-qubit MFM measurements did not lead to substantive improvements in the \metricId{} value for reconstructions using $K_{\binom{Q}{1} + \binom{Q}{2}}$ for \VAL, \YORK, and \CAM{} ($C_0, T_2$).  We note that the addition of spectator qubits can also lead to a reconstructed MFM with much higher values of \metricId{}, as in the case of \JOJO{} ($J_8$).  The effects of spectator qubits will be discussed further in Sec. \ref{sec:lambda2}

On the other hand, cluster MFMs measured on $>2$ qubit contain higher degrees of qubit correlations and we observe that this method leads to accurate reconstructions ($\Vert \Delta_K \Tilde{K} \Vert_F<0.1$) for \VAL, \YORK, and \CAM{} ($C_0$).

\subsection{SCF-based identification of qubit correlations}
\label{sec:lambda2}
In Section \ref{sec:cluster_MFM} we reported results on $\Lambda_{AB}$ as a measure of correlation between qubit subsets. In this section we calculate the SCF between individual pairs of qubits in a $n$-qubit layout.  In the absence of correlations all entries of $\Lambda_{ab}$ would be zero, but statistical noise can result in non-zero matrix values.  Using the uncertainty propagation from Sec. \ref{sec:uncertainty_prop} we state that two regions are correlated if the SCF satisfies: $\Lambda_{AB} > \sigma(\Lambda_{AB})$.  However, as mentioned in the above section, 1- and 2-qubit MFMs directly measured can only capture 1- and 2-qubit correlations. Combining the $\Lambda_{AB}$ matrix with spectator and extracted kernel methods outlined in \ref{sec:spectators_and_extracted} allows us to identify correlated regions and qubits on a chip.
\begin{figure}[htbp]
  \centering
  \includegraphics[width=\columnwidth]{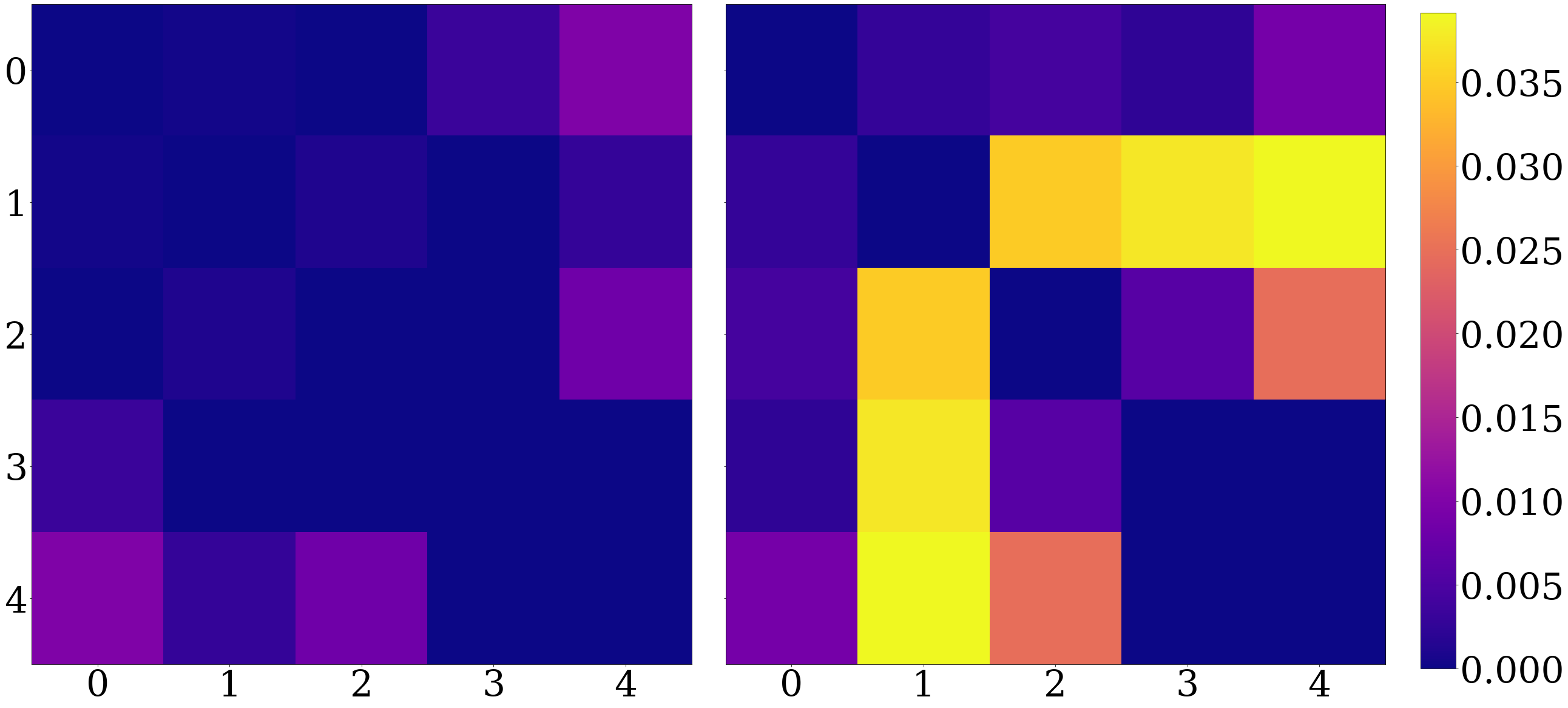}
  \caption{$[\Lambda_2 - \sigma(\Lambda_2)]$ matrix for \VAL: (Left) $1$- and $2$-qubit MFMs measured without spectator qubits, (Right) $1$- and $2$-qubit MFMs measured with spectator qubits.}
  \label{fig:Valencia_heatmap_spectators}
\end{figure} 
\begin{figure}[htbp]
  \centering
  \includegraphics[width=0.9\columnwidth]{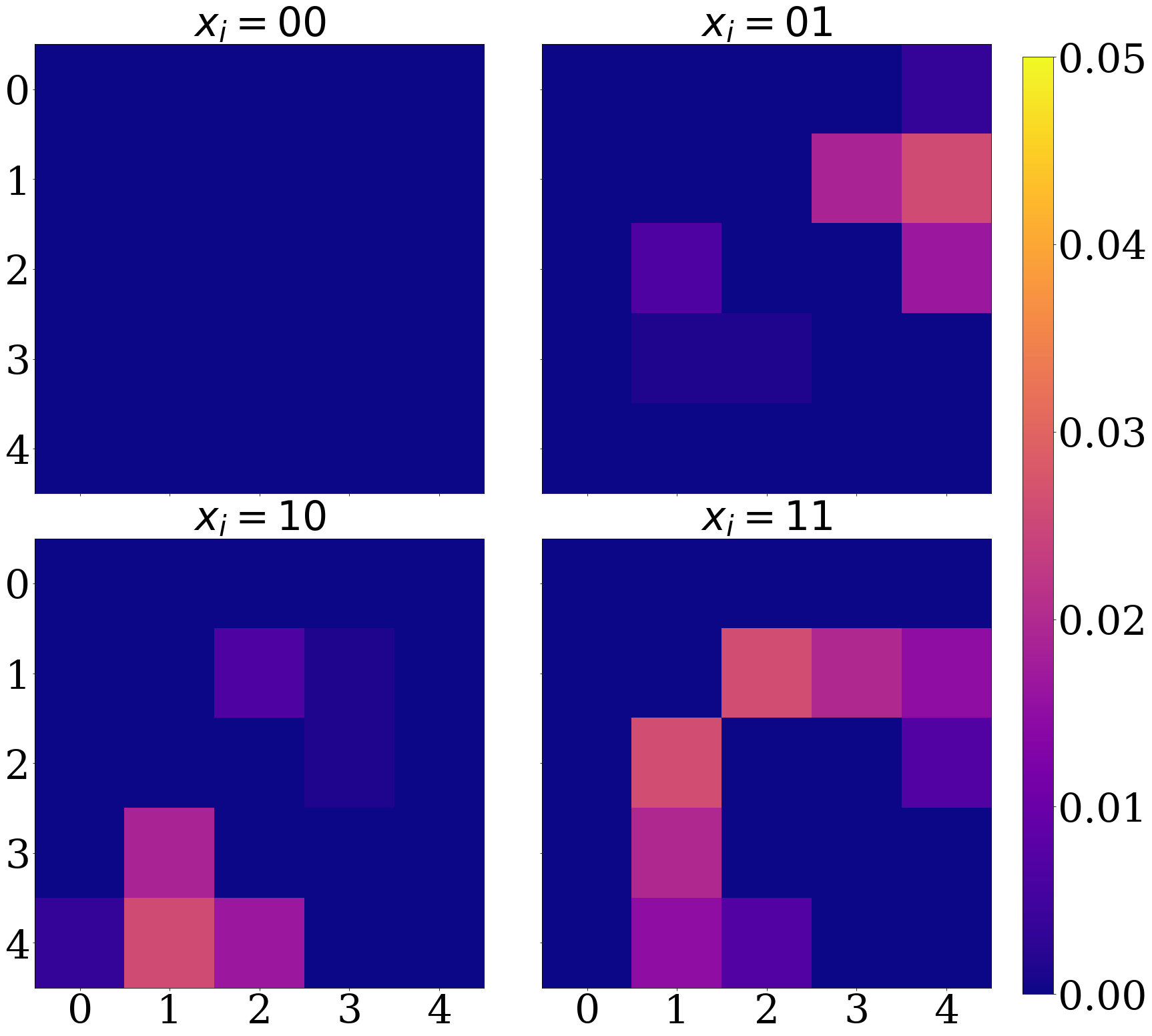}
  \caption{$[\Lambda_2(x_i) - \sigma(\Lambda_2(x_i)]$ matrices for $1$- and $2$-qubit MFMs measured with spectator qubits on \VAL. The plots show the correlations present when preparing specific target states $[00],[01],[ 10 ],[11]$.}
  \label{fig:Valencia_heatmap_states}
\end{figure} 
\begin{figure*}[htbp]
  \centering
  \includegraphics[width=0.8\textwidth]{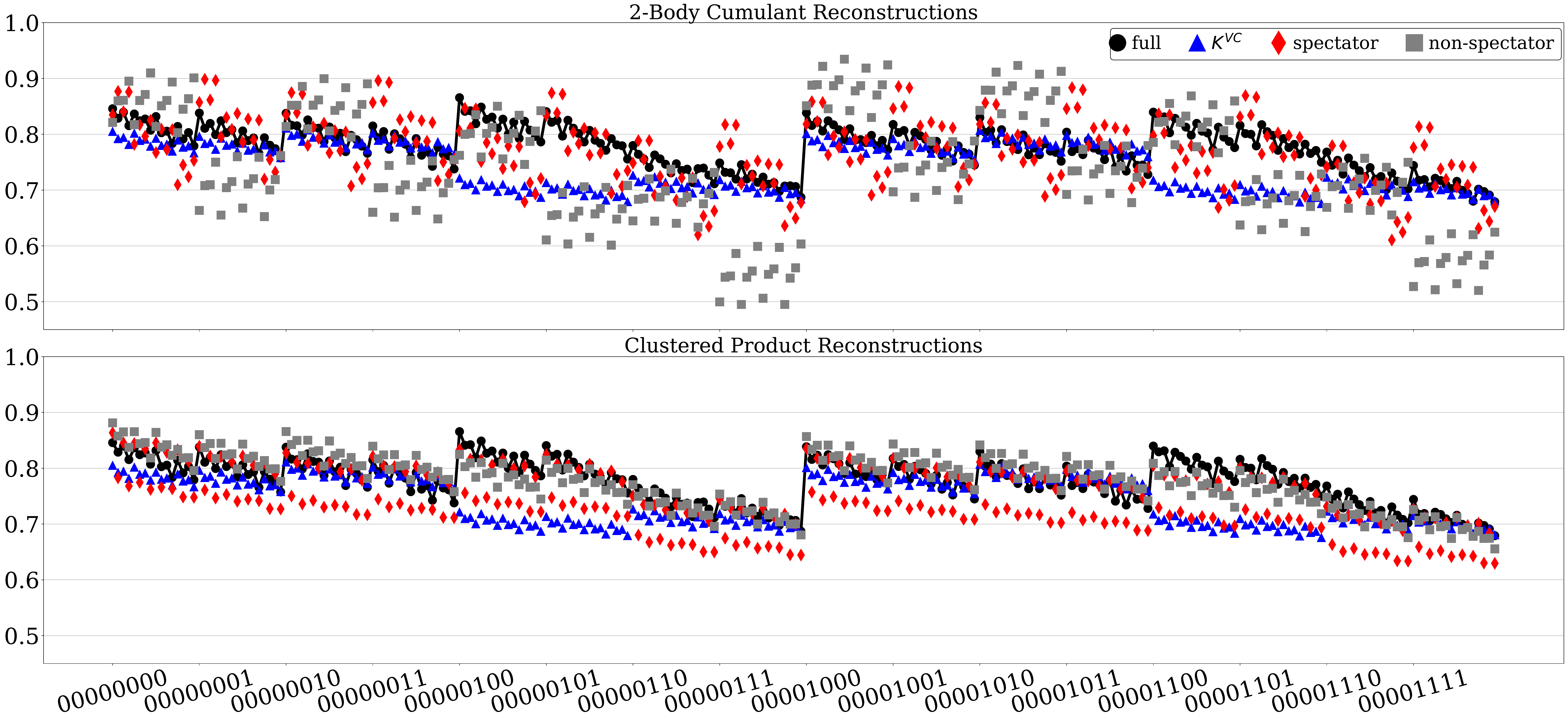}
  \caption{Reconstructed state fidelities measured on \JOJO{}. (Top) Reconstruction of 8-qubit MFM using $K^{VC}$ (blue, triangles), 2-body cumulant measured with spectators (red, diamonds), 2-body cumulant measured without spectators (grey, squares).  (Bottom) Reconstruction of 8-qubit MFM using $K^{VC}$ (blue, triangles), 4-qubit clusters measured with spectators (red, diamonds), 4-qubit clusters measured without spectators (grey, squares).}
  \label{fig:big_fidelity_plot}
\end{figure*} 

We use MFMs measured on \VAL{} as an example and compute $\Lambda_{AB}$ for all qubit pairs $a,b$.  As in Section \ref{sec:cluster_MFM}: we state that two qubits are correlated if the SCF satisfies: $\Lambda_{AB} > \sigma(\Lambda_{AB})$. 
In Fig. \ref{fig:Valencia_heatmap_spectators} we plot the statistically significant degree of correlation $|\Lambda_{ab} - \sigma(\Lambda_{ab})|$.

Directly measuring the $1$ and $2$ qubit MFM terms used in Eq. \ref{eq:two_body_term} shows weak correlations between qubit 0 and 4, and qubits 2 and 4.  However from the results reported in Table \ref{tab:TABLE1_cumulant_nospec_5qubit_chip} the $5$-qubit MFM constructed with only $1$ and $2$ qubit MFMs did not agree with the directly measured $5$-qubit MFM.  If the $1$ and $2$-qubit MFMs are extracted from measurements made with spectator qubits then we see that the $[\Lambda_2 - \sigma(\Lambda_2)]$ matrix shows a strong correlation between qubits $\{1, 2, 3, 4\}$.  In Fig. \ref{fig:metricId_5q_cluster} the (1,4) cluster MFM is formed by measuring a $4$-qubit MFM on $\{ 1,2,3,4 \}$ and the single hardware qubit $0$.  This reconstruction had the lowest SCF value and closest agreement between reconstructed MFM and full $5$-qubit MFM on \VAL.  From this we conclude that the 1 and 2-qubit MFMs are effective in identifying qubit-qubit correlations when spectator qubits are included.  

The values of $\Lambda_2$ (as plotted in Fig. \ref{fig:Valencia_heatmap_spectators}) sum over all terms in the two-body cumulant matrix; we can also decompose $\Lambda_2$ based on the target state $x_i$ to investigate the prevalence of qubit correlations with varying levels of gate noise.  From the results plotted in Fig. \ref{fig:Valencia_heatmap_spectators}[Right panel] we compute the $\Lambda_2$ matrix individually for each target state $[00],[01],[ 10 ],[11]$ and plot the results in Fig. \ref{fig:Valencia_heatmap_states}.  The correlations between $\{ 1, 2, 3, 4\}$ are not seen when preparing the state $x_i = 00$ (no gate errors) but become significant when preparing the state $x_i = 11$ (X and H gate errors prevalent).

In this work we have presented multiple methods for constructing MFMs.  On \JOJO{} using the hardware qubits $\{6,5,10,11,8,9,13,14\}$ we have computed the 8 qubit MFM using each method.  In Tables \ref{tab:TABLE2_5qubit_cumulant_subsets} and \ref{tab:TABLE3_cumulant_spec_5qubit_8qubit} we report the \metricId{}, \metricK{} and \metricF{} values for cumulant-based constructions. While the addition of spectators in the cumulant construction does not improve the \metricId{} and \metricK{} values, we observe different effects in the \metricF{} values.  In Fig. \ref{fig:big_fidelity_plot} we plot the diagonal elements of an 8 qubit MFM constructed in several ways: the full measured MFM, the cumulant-based reconstruction with and without spectators, and the cluster product reconstruction with and without spectators. The inclusion of spectator qubits in the measurement of each 4-qubit cluster improves the final state fidelities, compared to measurements of each 4-qubit cluster without spectators.  On the other hand, the inclusion of spectators in the measurements of 1 and 2-qubit MFMs used in the cumulant reconstruction showed no clear improvement compared to measurements without spectators. 

\section{Conclusions}
\label{sec:conclusions}
In this work we have introduced a matrix-based characterization method of qubit correlations.  Using this method as a high-level metric we can identify qubit correlations a quantum device: both the location (between which hardware qubits) and the strength (magnitude).  There are multiple sources of qubit correlations and at the level of our approach, we can distinguish between: statistical noise, measurement noise, and gate-level noise.  However fully characterizing the exact source of qubit-qubit correlations remains work for a future study. 

One promising application of this work is the concept of executing multiple circuits on a single QPU with the goal of preparing uncorrelated states on subsets of qubits (e.g. \cite{PhysRevApplied.11.034021}).  This spatial multiplexing of quantum states is analogous to multiplexing used in digital communication to send multiple messages.  The viability of this application depends on identifying qubit subsets that have minimal correlated noise between them.  The cumulant analysis introduced in this paper can be used to find highly correlated qubit subsets on a QPU to avoid using them in experiments to minimize correlated error. In practice, this data can be measured independent of constructing a specific MFM. We emphasize that the results presented here gauge the independence between qubits based on the independence of their projected measurements in the computational basis.  The $\lambda_2$ matrix introduced in this work supports the further development of multiplexing as a means to increase the efficiency of NISQ-era quantum computing.

\section*{Acknowledgements}
The authors would like to thank Vicente Leyton-Ortega for insightful discussions about noise characterization and error mitigation.  The authors would like to thank IBM for providing information about the IBM Q system.  

This work was supported as part of the ASCR Quantum Testbed Pathfinder Program at Oak Ridge National Laboratory under FWP \#ERKJ332.  This research used quantum computing system resources of the Oak Ridge Leadership Computing Facility, which is a DOE Office of Science User Facility supported under Contract DE-AC05-00OR22725. Oak Ridge National Laboratory manages access to the IBM Q System as part of the IBM Q Network.

\section*{Conflict of interest}
The authors declare that they have no conflict of interest.
 
\begin{appendices}
\section{XACC Implementation}
\label{appendix:xacc_implementation}

The eXtreme-scale ACCelerator programming framework (XACC) provides a system-level, 
quantum-classical software infrastructure that is extensible, modular, and hardware agnostic \cite{mccaskey2020xacc}. 
XACC enables the programmability of quantum kernels that may be executed in a uniform manner 
on backends provided by IBM, Rigetti, D-Wave, IonQ, and a number of numerical simulators. Moreover, 
there exists extensive support for typical hybrid variational workflows through a uniform 
\texttt{Algorithm} interface. 

For this work, we have leveraged this framework and contributed a unique implementation 
of the \texttt{Algorithm} interface specifically enabling the data-driven circuit learning 
workflow in an extensible and modular manner. Our implementation remains general with 
regards to the circuit ansatz, the backend quantum computer targeted, the loss function, 
and gradient computation strategy. Moreover, XACC provides extensibility with regards to 
typical error mitigation strategies that apply general pre- and post-processing of 
quantum execution results. XACC provides this through standard object-oriented decoration 
of the provided backend \cite{gof}. For this work, we have extended this mechanism with support 
for automated error kernel generation and mitigation. 

Fig. \ref{fig:ddcl_xacc} demonstrates how one might leverage the error mitigation strategies 
described in this work. We provide a simple two qubit example whereby we run the 
typical data-driven circuit learning workflow in a manner that is automatically error 
mitigated. Programmers begin by specifying the desired \texttt{Accelerator} backend, here we select \BOBO, 
the IBM Boeblingen machine. We then allocate four qubits using the C-like quantum malloc, \texttt{qalloc}. This specifies we 
are running on a four qubit register and the \texttt{qbits} variable will later serve to 
hold the computation results as well as additional information about the computation. 
After specifying the backend and the qubit buffer, we now decorate the backend 
to endow the execution with the specified error mitigation strategy. The command line arguments , 'gen-kernel' and 'layout' specify 
whether one would like to generate a new error kernel on this run or use one from a previous 
run, and layout determines the mapping from logical qubits to physical qubits. We then specify 
the circuit we wish to execute. The accelerator decorator will then invoke the accelerator 
and generate the error kernel on the specified backend from a subroutine defined in the 
accelerator decorator. Then, the resultant distribution is corrected according to the 
inverse of the kernel and the new distribution is returned. 
\begin{figure}[t!] 
\begin{python}
import xacc
# Get the QPU and allocate a single qubit
qpu = xacc.getAccelerator('ibm:ibmq_boeblingen')
qbits = xacc.qalloc(2)
layout = [2, 4]

# Decorate the QPU with the assignment-error-kernel 
# error mitigation strategy 
qpu = xacc.getAcceleratorDecorator(
                'assignment-error-kernel', qpu,
                {'genKernel':True,
                 'layout':layout
                })
                
# Get the MLPack Optimizer, default is Adam
optimizer = xacc.getOptimizer('mlpack')

# Create a simple quantum program
xacc.qasm('''
.compiler xasm
.circuit state_prep
.parameters x
.qbit q
U(q[0], x[0], -pi/2, pi/2);
U(q[0], 0, 0, x[1]);
U(q[1], x[2], -pi/2, pi/2);
U(q[1], 0, 0, x[3]);
CNOT(q[0], q[1]);
U(q[0], 0, 0, x[4]);
U(q[0], x[5], -pi/2, pi/2);
U(q[1], 0, 0, x[6]);
U(q[1], x[7], -pi/2, pi/2);
''')

# Get the circuit, specify physical qubits
f = xacc.getCompiled('state_prep')
f.defaultPlacement(qpu, {'qubit-map':layout})

# Get the DDCL Algorithm,
ddcl = xacc.getAlgorithm('ddcl', 
            {'ansatz': f,
            'accelerator': qpu,
            'target_dist': [.5,.5],
            'optimizer': optimizer,
            'loss': 'js',
            'gradient': 'js-parameter-shift'
            })
# execute
ddcl.execute(qbits)
# Print the error-mitigated result
print('Error Mitigated Optimal Loss = ', 
        qbits['opt-val'])
\end{python}
\caption{Code snippet demonstrating the data-driven circuit learning workflow leveraging automated error mitigation.}
\label{fig:ddcl_xacc}
\end{figure}
\end{appendices}
\bibliographystyle{unsrt}

\end{document}